\documentclass[9pt,twocolumn,twoside,lineno]{pnas-new}
\templatetype{pnasresearcharticle}
\title{Tunable Bi-frustrated Electron Spin and Charge States in a Cu-Hexaaminobenzene Framework}
\author[a]{Wei Jiang}
\author[b,1]{Zheng Liu}
\author[c,a,1]{Jia-Wei Mei}
\author[d,a]{Bin Cui}
\author[e,a,1]{Feng Liu}

\affil[a]{Department of Materials Science and Engineering, University of Utah, Salt Lake City, UT 84112, USA}
\affil[b]{Institute for Advanced Study, Tsinghua University, Beijing 100084, China}
\affil[c]{Institute for Quantum Science and Engineering, and Department of Physics, Southern University of Science and Technology, Shenzhen 518055, China}
\affil[d]{School of Physics, Shandong University, Jinan 250100, China}
\affil[e]{Collaborative Innovation Center of Quantum Matter, Beijing 100084, China}

\leadauthor{Jiang}

\significancestatement{Both electron spin and charge can get frustrated in a geometrically frustrated lattice. If spins are favored to align antiferromagnetic in up and down direction on neighboring sites, they will be frustrated in a triangle lattice; if electrons move too slow in a lattice with a narrow bandwidth of quenched kinetic energy, they are in a frustrated charge state incapable of hopping independently because of Coulomb repulsion. Such frustrations leading to exotic quantum phases have fascinated theorists for a long time. However, real materials with frustrated spin lattice have only been identified recently. Here, we discover that a newly synthesized metal-organic framework [Cu$_3$(C$_6$N$_6$H$_6$)$_2$] hosts simultaneously both frustrated spins and electrons, affording an exciting platform to explore new exotic quantum phases.
}

\authorcontributions{W.J., Z.L., J.-W.M., and F.L. designed research; W.J. and J.-W.M. performed research; B.C. contributed new analytic tools; W.J., Z.L., J.-W.M., and F.L. wrote the paper. }
\authordeclaration{The authors declare no conflict of interest.}
\correspondingauthor{\textsuperscript{1}To whom correspondence should be addressed. E-mail: fliu@eng.utah.edu, meijw@sustc.edu.cn, zheng-liu@mail.tsinghua.edu.cn.}


\begin{abstract}
A geometrically frustrated lattice may host frustrated electron spin or charge states that spawn exotic quantum phases. We show that a newly synthesized metal-organic framework of Cu-Hexaaminobenzene [Cu$_3$(HAB)$_2$] exhibits a multi spectra of unusual quantum phases long sought after in condensed matter physics. On one hand, the Cu$^{2+}$ ions form an ideal $S$-1/2 antiferromagnetic kagome lattice. On the other hand, the conjugated-electrons from the organic ligands give rise to completely dispersionless energy bands around the Fermi level, reproducing a frustrated $\pi_x$-$\pi_y$ hopping model on a honeycomb lattice. We propose to characterize the coexistence of frustrated local spins and conjugated electrons through scanning tunneling microscopy simulations. Most remarkably, their close energy proximity enables one to tune the system between the two frustrated states by doping up to one hole per HAB unit. Thus, Cu$_3$(HAB)$_2$ provides a unique exciting platform to investigate the interplay of frustrated spins and electrons in one single lattice, e.g. by gating experiments, which will undoubtedly raise interesting theoretical questions leading to possible new condensed-matter phases.

\end{abstract}
\dates{This manuscript was compiled on \today}
\doi{\url{www.pnas.org/cgi/doi/10.1073/pnas.XXXXXXXXXX}}
\begin{document}
\verticaladjustment{-2pt}

\maketitle
\thispagestyle{firststyle}
\ifthenelse{\boolean{shortarticle}}{\ifthenelse{\boolean{singlecolumn}}{\abscontentformatted}{\abscontent}}{}

\begin{figure*}
  \centering
\includegraphics[width=15cm]{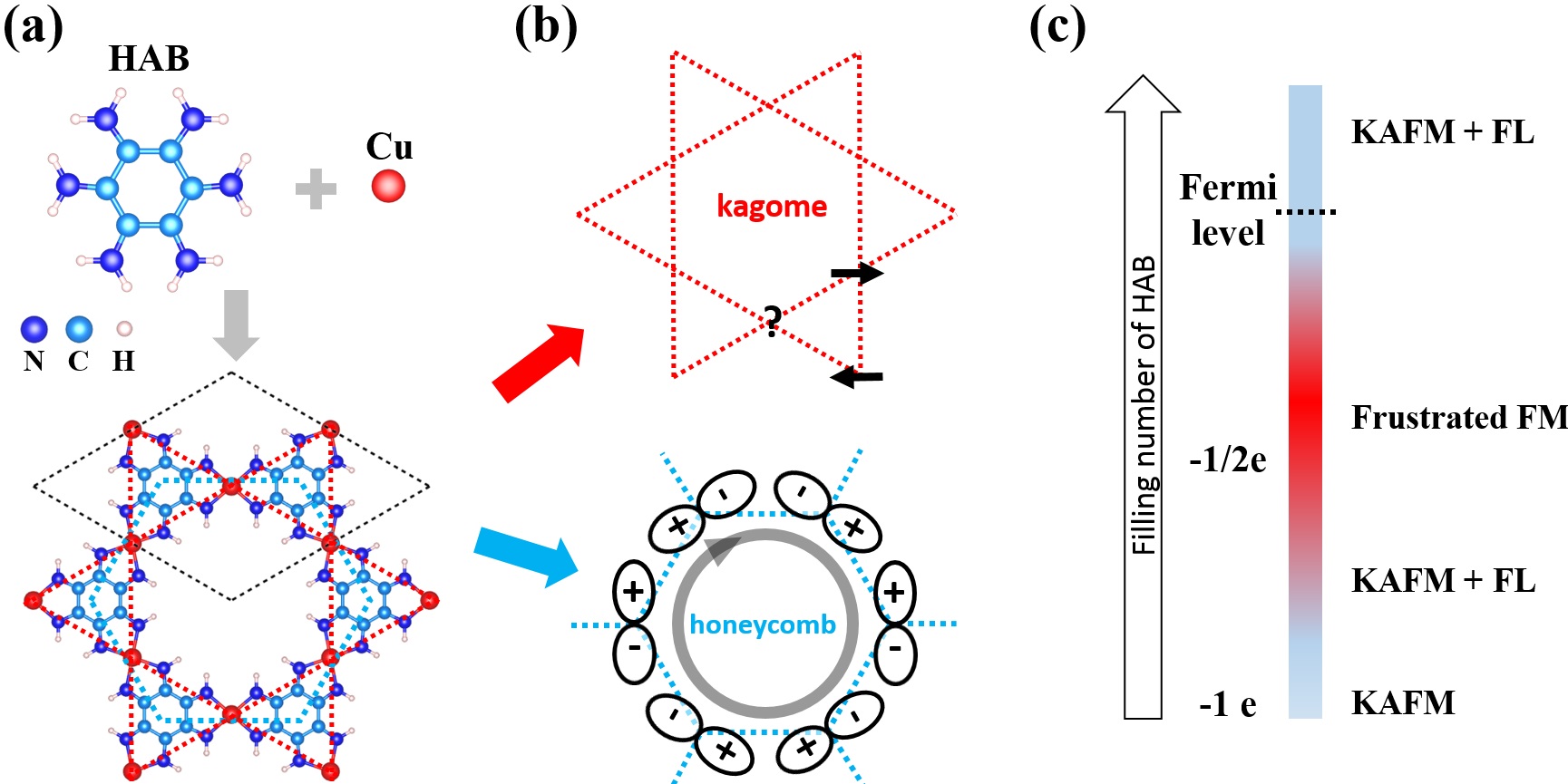}
\caption{Structural and electronic properties of Cu$_3$(HAB)$_2$. (a) Atomic structure of a HAB molecule and the crystalline structure of Cu$_3$(HAB)$_2$ by coordinating HAB with Cu. (b) Schematics of bifrustration in this material. (c) Phase evolution as a function of electron filling number of HAB.}
\label{fig:figure1}\end{figure*}

\dropcap{D}evelopment in modern chemical synthesis is turning metal-organic frameworks (MOF) into a playful legoland for condensed matter physicists, where lattice structures with desired electronic and magnetic properties can be delicately tailored with diverse metal ions and organic linkers~\cite{Lu2014,Deria2014,Brozek2014,Colon2014}. The most recent fascination is about the ``frustrated'' spins or electrons triggered by special lattice geometry. The frustration is considered to suppress classical orders and lead to highly nontrivial quantum-entangled phases, such as quantum spin liquid (QSL) state\cite{Anderson1987,Lee2008,Balents2010}. Experiments in inorganic systems, namely herbertsmithite~\cite{Shores2005,Han2012,Fu2015} and Zn-barlowite~\cite{Feng2017}, suggest a gapped QSL in a kagome spin system although the existence of the gap is still under debate~\cite{Yan2011,Depenbrock2012,Jiang2012,Mei2017,He2016,Jiang2016,Liao2016}. Several organic molecular crystals are among the most promising QSL candidates~\cite{Shimizu2003,Kurosaki2005,Itou2007,Itou2008,Isono2014}, e.g. $\kappa$-(BEDT-TTF)$_2$Cu$_2$(CN)$_3$ and EtMe$_3$Sb[Pd(dmit)$_2$]$_2$, both of which are found to have a charge insulating yet spin fluctuating ground state. Very recently, it was proposed that Kitaev spin liquid could be realized in a MOF with Ru$^{3+}$~\cite{Yamada2017}.

There are also first-principles MOF designs of ``frustrated metal''. For example, in an In-phenylene framework~\cite{Liu2013}, calculation predicts completely dispersionless electronic bands around the Fermi level. The underlying physics is described by a frustrated $p_x$-$p_y$ hopping model on a honeycomb lattice, which was previously mainly studied in cold-atom setups~\cite{Wu2007,Wu2008}. Because the kinetic energy is largely quenched, the dispersionless band, or ``flat band'', naturally becomes an interaction-dominated platform, giving rise to various instabilities, such as charge-density wave, ferromagnetism, superconductivity and even high-temperature fractional quantum Hall states~\cite{Tang2011,Sun2011,Neupert2011,Li2014a}.

Here, we discover a rare example, i.e. a Cu-hexaaminobenzene framework [Cu$_3$(HAB)$_2$, Cu$_3$(C$_6$N$_6$H$_6$)$_2$], that simultaneously hosts both frustrated spins and frustrated electrons. The crystalline Cu$_3$(HAB)$_2$ has already been synthesized by coordinating hexaaminobenzene (HAB) ligands and Cu ions through air-liquid or liquid-liquid interfacial interaction~\cite{Lahiri2017}. As schematically shown in Fig.~\ref{fig:figure1}(a), the HAB ligands connect Cu ions into a two-dimensional (2D) periodic lattice; each Cu ion locally coordinates with four N atoms in a distorted planar square geometry. We reveal that this material can be viewed as a combination of two subsystems [Fig.~\ref{fig:figure1}(b)]. One is a kagome antiferromagentic $S=1/2$ lattice (KAFM) consisting of the Cu$^{2+}$ ions. KAFM is considered as one of the most promising systems to realize QSL\cite{Norman2016}. The most fascinating feature in KAFM is the possible existence of fractional spinon excitations, which was speculated in theory decades ago and has been recently observed in experiments~\cite{Han2012,Feng2017}. The other subsystem consists of $p_{x,y}$-like ($\pi_{x,y}$) molecular orbitals on a honeycomb lattice, which give rise to an itinerant flat band right at the Fermi level. Because the flat band is susceptible to various spontaneous symmetry breaking, electrons in this band are considered to be in a fragile Fermi liquid (FL) state. Remarkably, the two subsystems are strongly coupled within a close energy proximity, leading to the possibility of new phases with different amount of doping, which is experimentally feasible. The system can be readily tuned, e.g. by gating experiments into a series of quantum states with different electron filling, as illustrated in Fig.~\ref{fig:figure1}(c). The evolution of these quantum states will be discussed qualitatively before the end of this Article. The coupling between these two subsystems poses an important but difficult theoretical question, and a complete answer may strongly rely on new experimental input. The primary goal of the present work is to computationally substantiate the proposed bifrustration in this real material, and to establish a basis for future investigation.

\section*{Calculation results}

\begin{figure}[tbp]
  \centering
\includegraphics[width=\columnwidth]{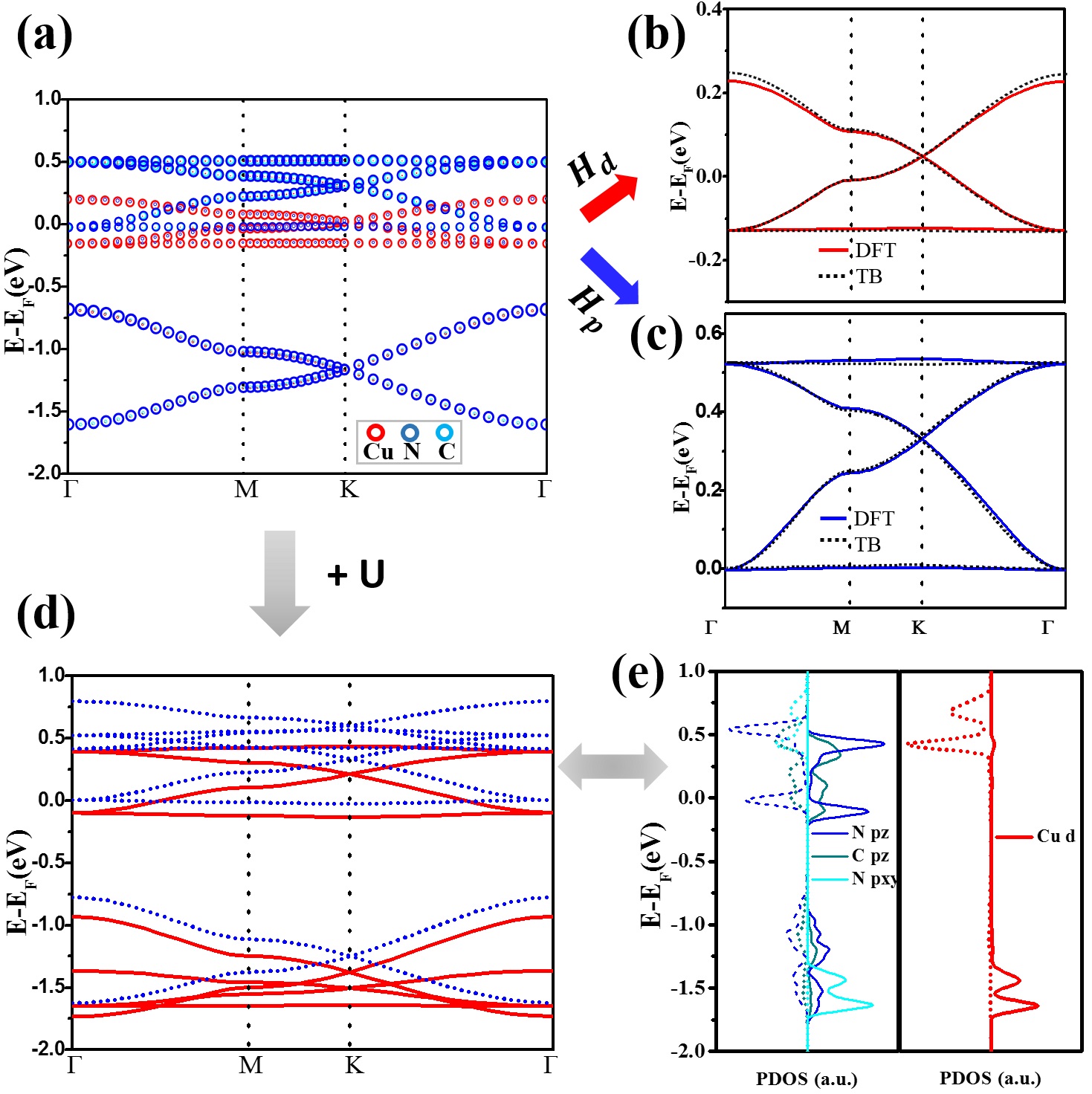}
\caption{Band structures from DFT, DFT+U and the effective TB models. (a) Atomic-resolved DFT band structure; the marker size indicates the weight of the atomic composition. Decomposition of the bands into (b) a kagome lattice and (c) a honecycomb lattice, with comparison to the effective TB model band structure.(d) The DFT+U band structure; the solid(red) and dashed(blue) curves corresponds to different spins. (e) The DFT+U projected density of states.}
\label{fig:figure2}\end{figure}

\begin{figure}[tbp]
\centering
\includegraphics[width=\columnwidth]{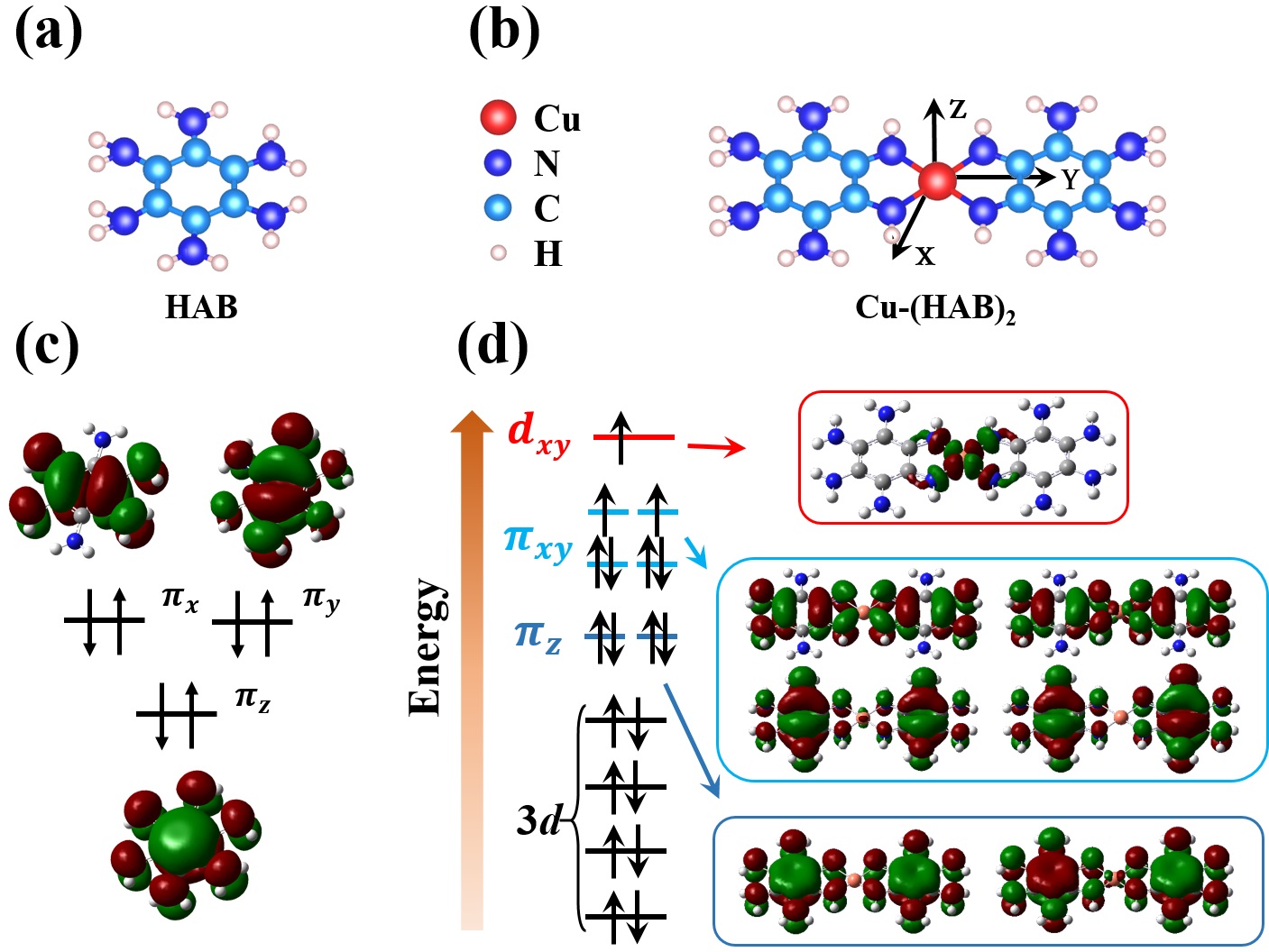}
\caption{Energy levels and the associated wavefunction of (a) a HAB molecule and (b) a Cu-(HAB)$_2$ cluster }
\label{fig:figure3}\end{figure}

We first calculated the band structure within the standard density functional theory (DFT, Supporting Information) [Fig.~\ref{fig:figure2}(a)]. Below the Fermi level (from -1.75 to -0.5 eV), there is a pair of isolated $\pi_z$ Dirac bands just like the $p_z$-bands in graphene. Around the Fermi level (from -0.25 to 0.5 eV), there are seven bands, among which three are completely dispersionless, manifesting typical features of frustration. More importantly, the Fermi level almost coincides with one of the flat band. By resolving these bands according to the atomic composition, they can be decomposed into a three-band set (red) and a four-band set (blue). The three-band set mainly arises from the Cu ions, which can be nicely fitted by a single-orbital tight-binding (TB) hopping model on a kagome lattice [Fig.~\ref{fig:figure2}(b)]:
\begin{eqnarray}
\label{eq:Kagome}
  H_d^0=t_d\sum_{\langle ij\rangle}(d_{i\sigma}^\dag d_{j\sigma} + \text{H.c.})
\end{eqnarray}
where the summation $\langle ij\rangle$ are constrained within the nearest-neighbors (NN) on the Cu kagome lattice indexed by $i,j$, and $t_d=0.06$ eV is the fitted NN hopping amplitude. The four-band set arises from a combination of N and C orbitals, which reflects the typical features of an effective $\pi_{xy}$ $\sigma$-bond hopping model on a honeycomb lattice:
\begin{equation}
\begin{split}
\label{eq:pxy}
H_{\pi}^0 = t_\pi\sum_{\langle IJ\rangle}(\pi_{I}^{l\dag}\pi_{J}^l+ \text{H.c.})
\end{split}
\end{equation}
where the summation $\langle IJ\rangle$ are constrained within the NN bonds on the HAB honeycomb lattice indexed by $I,J$. $\pi^{l\dag}$ creates a superposition state of $\pi_x$ and $\pi_y$ orbitals with the orbital vector parallel to the bond, i.e. $|\pi^l\rangle=\cos\phi|\pi_x\rangle+\sin\phi|\pi_y \rangle$ with $\phi$ as the angle between the bond and the x-axis. $\pi_{x,y}$ are molecular conjugated orbitals from HAB with analogical symmetries to $p_{x,y}$ orbitals in single C or N atom. According to Fig.~\ref{fig:figure3}(a), they distribute among both C and N atoms. In this model, $t_\pi$ is equivalent to the $\sigma$-bond hopping amplitude, while the weak $\pi$-bond hopping is neglected. The fitted value for $t_\pi$ is 0.22 eV [Fig.~\ref{fig:figure2}(c)]. We noticed that the LDA band structure could be better reproduced by further considering that the $\pi^l$-orbitals on the NN sites have a small overlap $S=\langle \pi^l_I|\pi^l_J \rangle=0.18$. The generalized eigenvalue problem becomes $\det(H_\pi^0-\epsilon S)=0$, where $\epsilon$ is the eigenenergy.

To fully characterize this intriguing material beyond DFT, two additional questions should be answered: (a) What is the basis of these hopping models? The nonorthogonality of the effective $\pi_{x,y}$-orbitals revealed from the band fitting already implies that they should be complicated molecular orbitals; (b) How does interaction change the picture? This question is particularly important to the kagome bands arising from Cu, because Cu 3$d$-electrons are known to experience strong Coulomb repulsion.

These questions can be partially addressed by studying the building blocks of Cu$_3$(HAB)$_2$ based on the state-of-the-art quantum chemistry computational package (Gaussian). We proceed by a ``computational thought experiment'' to first analyze a single HAB molecule [Fig.~\ref{fig:figure3}(a)] and then sandwich one Cu atom between two HAB molecules [Fig.~\ref{fig:figure3}(b)]. Figure~\ref{fig:figure3}(c) shows the three highest occupied levels of a single HAB molecule. We also plot the wavefunction of these three levels. It is clear that they mainly arise from the $\pi$-conjugation of the C/N $p_z$-orbitals. The two degenerate levels $\pi_x$ and $\pi_y$ on top have the same symmetry as the atomic $p_x$ and $p_y$ orbitals. The single level below $\pi_z$ has the same symmetry as the atomic $p_z$ orbital. The molecular  $\pi_{x,y}$-orbitals define the basis of Eq.~\ref{eq:pxy}. In a single molecule, $\pi_{x,y,z}$  are all fully occupied.

Figure~\ref{fig:figure3}(d) shows the evolution of the occupied levels after putting a Cu atom between two HAB molecules. The highest occupied level now becomes a half-filled single orbital primarily consisting of the Cu 3d orbital. By choosing the Cu-N bonds as the orbital quantization axes, this molecular orbital is equivalent to an atomic $d_{x^2-y^2}$-orbital. All the other Cu 3$d$-orbitals are found to be deep down away from the Fermi level and fully occupied. Hence, Cu is in a +2 (3$d^9$) valence state, or equivalently, contains one single hole. The situation is actually very similar to that in high-T$_c$ cuprate superconductors\cite{Lee2006}, where the Cu-O planar crystal field splits out a half-filled $d_{x^2-y^2}$-orbital, and the dynamics of this dangling bond is the microscopic origin of all the unusual electronic properties\cite{Lee2006}. The $\pi_{x,y}$ orbitals now become the second highest occupied levels. They are still degenerate, but the filling factor reduces from 1 to $\frac{3}{4}$.This result can be understood by noticing that to bond with Cu, the two HAB molecules release four H$^{+}$ ions in total. The Cu$^{2+}$ ion, however, can only donate two electrons to the HAB molecules. Consequently, upon bonding with a Cu, each HAB molecule loses one out of its four $\tilde{p}_{x,y}$ electrons.

We can now re-evaluate the DFT results. The effective four-band model on a honeycomb lattice (Eq.~\ref{eq:pxy}) can be firmly constructed out of the $\pi_{x,y}$ orbitals. These orbitals are itinerant, and thus should be well captured by DFT. The filling factor of the four-band set should be $\frac{1}{4}$, considering that each HAB ligand bonds with three Cu ions in the periodic Cu$_3$(HAB)$_2$ framework (Supporting Information). This means that the bottom flat band is fully occupied and all the other three bands are empty. The location of the Fermi level calculated from LDA correctly captures this property. The origin of the flat band lies in a special localized eigenstate as sketched in Fig. ~\ref{fig:figure1}(b). Electrons in this eigenstate cannot leak outside the hexagon due to complete destructive interference~\cite{Wu2007}.

With respect to the kagome degree of freedom, the Cu-(HAB)$_2$ cluster calculation indicates that each Cu$^{2+}$ ion contains a local spin, whereas the DFT calculation suggests the formation of a nonmagnetic metal with the Fermi level crossing the half-filling energy. This is the widely known failure of DFT due to the underestimation of electron correlation. A simple remedy within the DFT formalism is to include electron correlations in a Hartree-Fock mean-field manner, know as the +U correction. It has been demonstrated that DFT+U can correctly reproduce Mott insulating gap by explicitly breaking the time-reversal symmetry. Figure~\ref{fig:figure2}(d) shows the resulted band structure by choosing the empirically averaged Coulomb interaction $\bar{U}\sim5$ eV as commonly used for Cu. By referring to the projected density of states (PDOS) [Fig.~\ref{fig:figure2}(e)], we found that the Cu kagome band is split into a set of lower Hubbard bands and a set of upper Hubbard bands. The Mott gap ($U_0$) is about 2 eV, much smaller than $\bar{U}\sim5$ eV. It reflects the strength of onsite Coulomb repulsion in the presence of Coulomb screening due to itinerant conjugated $\pi$-electrons in the system. The complete Hamiltonian for this Cu$^{2+}$ degree of freedom can then be revised into:
\begin{eqnarray}
\label{eq:Heisenberg}
H_d=H_d^0+U_0\sum_In_{i\uparrow}n_{i\downarrow}.
\end{eqnarray}
The hopping amplitude $t_d=0.06$ eV in $H_d^0$ is much smaller than $U_0$. Therefore, at half filling, the large-U perturbation can be applied, which maps $H_d$ into a $S$-1/2 Heisenberg model:
\begin{eqnarray}
\label{eq:KagomeU}
H_d\simeq J_d\sum_{\langle ij\rangle}\mathbf{S}_i\cdot\mathbf{S}_j,
\end{eqnarray}
where $J_d=\frac{4t_d^2}{U_0}\simeq7$ meV is the strength of the AFM NN spin exchange. This AFM superexchange strength is small because of the large distance between Cu$^{2+}$ ions. Furthermore, the $\pi_{x,y,z}$ molecular orbitals around the Fermi level primarily arise from the C/N atomic $p_z$ orbitals, which are orthogonal to the $d_{x^2-y^2}$-orbital. Thus, there is no charge transfer between $\pi_{x,y}$ orbitals on HAB ligands and $d_{x^2-y^2}$-orbital on Cu ions (Supporting Information). For the same reason, the superexchange for Cu 3$d$ orbitals is scaled by on-site Coulomb interaction $U_0$. 

One more important point revealed in Fig.~\ref{fig:figure2}(d) is that a Zeeman splitting of the $\pi_{xyz}$ bands occurs. This splitting is not related to the instability of the $\pi_{xy}$ flat band, because the $\pi_{z}$ Dirac bands are also split of the same amount. Therefore, it should be attributed to the coupling with the Cu local spin. It is clear from Fig.~\ref{fig:figure2}(e) that the downshifted (upshifted) $\pi_{xyz}$ bands have the same (opposite) spin direction as the Cu spin, indicating a ferromagentic (FM) Hund's coupling:
\begin{eqnarray}
\label{eq:hund}
H_{\pi d}=-J_H\sum_{i}\sum_{I\in \text{NN}(i)} \mathbf{S}_i\cdot \mathbf{s}_{I},
\end{eqnarray}
where $\mathbf{S}_i$ and $\mathbf{s}_{I}=\sum_{\alpha=x,y}^{\sigma=\uparrow/\downarrow}\sigma\pi_{I\alpha\sigma}^\dag\pi_{I\alpha\sigma}$ are the spin operators of the Cu $d_{x^2-y^2}$ electron and the NN HAB $\pi_{xy}$ electrons, respectively. $J_H$ denotes the Hund's coupling strength, which is roughly of the same magnitude as the Zeeman splitting $\sim$0.1 eV.

\begin{figure}[tbp]
  \centering
\includegraphics[width=0.9\columnwidth]{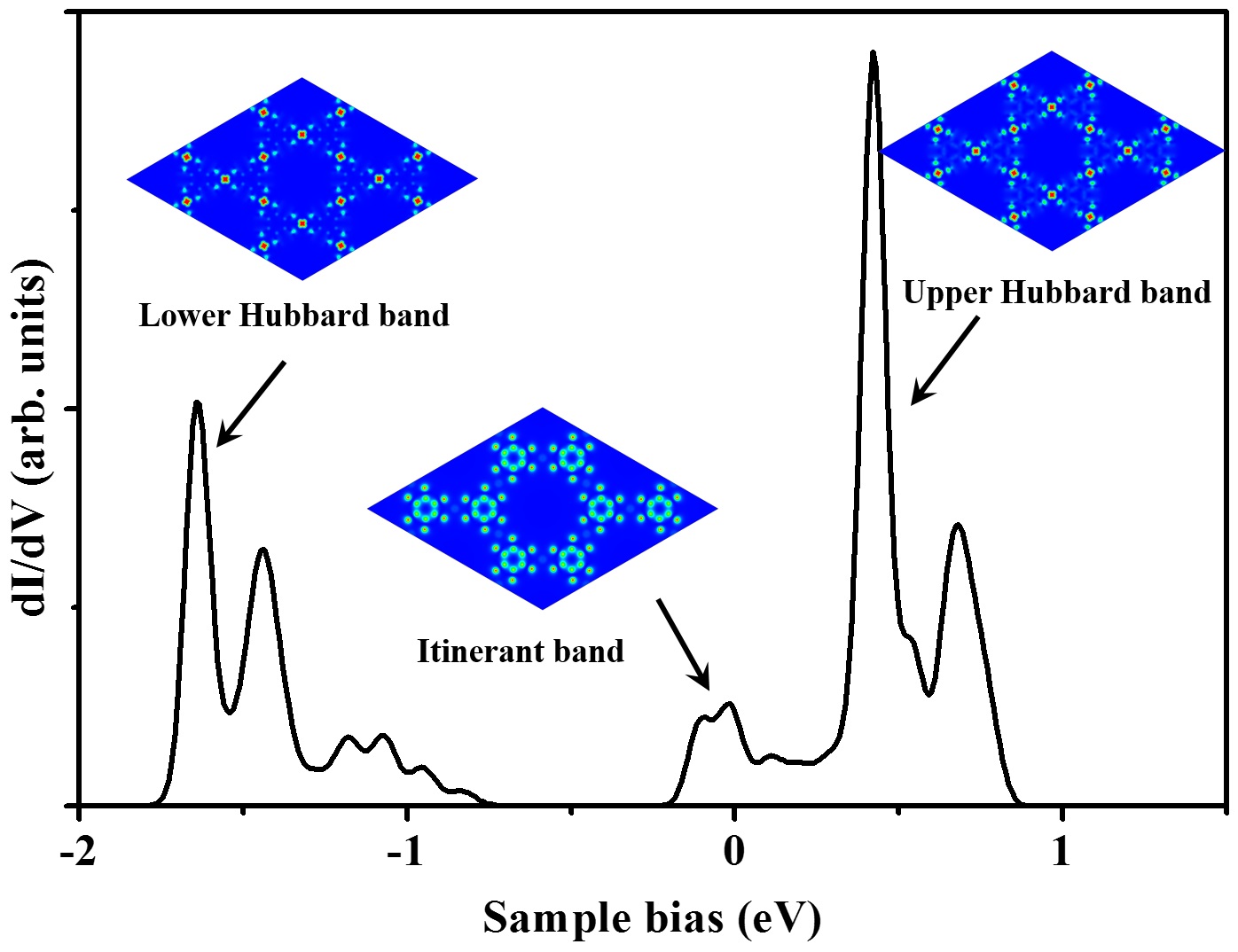}
\caption{STM simulations. Main figure displays the dI/dV spectra simulated for Cu$_3$(HAB)$_2$ surface, showing two prominent Hubbard peaks below and above the Fermi level along with one itinerant peak at the Fermi level. The insets show the STM topography of Cu$_3$(HAB)$_2$ surface with bias of 0.45/-1.55 and 0 eV corresponding to the kagome upper/lower Hubbard and honeycomb itinerant band, respectively.}
\label{fig:figure5}\end{figure}

To directly identify these two frustrated degrees of freedom, we propose to carry out scanning tunneling microscopy (STM) measurement of this material. STM has been widely applied to study the self-assembled metal-organic coordinated systems built through intermolecular hydrogen, covalent and coordinate bonding~\cite{Bartels2010,Liu2011}. Among those studies, many 2D network structures have been successfully synthesized through coordination bond between Cu and N atoms on various substrates~\cite{Lin2007,Pawin2008,Tait2008}, which makes growing and studying monolayer Cu$_3$(HAB)$_2$ through STM feasible and promising. Thus, we have simulated scanning tunneling spectroscopy (STS) mapping at two characteristic regions, as demonstrated in Fig.~\ref{fig:figure5}. The dI/dV spectra shows two prominent Hubbard peaks blow and above the Fermi level along with itinerant peaks right on the Fermi level. The STM topography simulations around the upper/lower hubbard peak with a bias of 0.45/-1.55 eV exhibit clear kagome patterns, and a honeycomb feature is observed for the bias of 0 eV around the itinerant peaks, as shown in the insets of Fig.~\ref{fig:figure5}. It is worth mentioning that our STM results are simulated for insulating substrates, where all the states will remain intact in the insulating gap.

\section*{Dicussion}

By combining the results above, we arrive at a full effective hamiltonian characterizing this unique material:
\begin{eqnarray}
\label{eq:full}
H=H^0_\pi+H_d+H_{\pi d},
\end{eqnarray}
which appears like a typical spin-fermion model as employed for iron pnictide high-temperature superconductors\cite{Kamihara2008,Dai2009,Kou2009} and colossal magnetoresistance manganites~\cite{Helmolt1993,Ramirez1997}. The interplay between the conducting electrons and local spins also lead to ubiquitous physical phenomena in other systems, such as the Kondo problems~\cite{Kondo1964,Anderson1961}, the heavy fermion systems~\cite{Andres1975,Senthil2004}, cuprate high temperature superconductors~\cite{Bednorz1986,Lee2006,Mei2012}. The situation of Cu$_3$(HAB)$_2$ is actually even more exotic - beyond the dichotomy between electron localization and itinerancy, as the spin and fermion also simultaneously suffer from strong frustration.

This bifrustrated spin-fermion model presents a new theoretical problem that remains to be explored by strong-correlation calculation methods, such as density matrix renormlization group~\cite{White1992}. On the other hand, the electron filling of the HAB ligands can be delicately tuned through either gating or chemical redox control~\cite{Kambe2013,Kambe2014}, which will result in various interesting quantum states.

Here, we will give some preliminary discussions about the phase evolution of Cu$_3$(HAB)$_2$ with different doping concentration, based on the existing knowledge of the subsystems. When all the $\pi$ electrons are removed (1e per HAB ligand, 1.24$\times$10$^{14}$ cm$^{-2}$), the system is at the $H_d$ side of the KAFM state, which is a hot topic in condensed matter physics in the past decade~\cite{Norman2016}. It is generally believed that $H_d$ alone will give rise to QSL with fractional spinon excitations. After adding $H^0_\pi$ by doping small amount of $\pi$ electrons, one simple possibility is that these fractional spinons can peacefully coexist with frustrated fermions in some circumstances, forming together a ``fractional Fermi liquid'' (FL$^*$)\cite{Senthil2003,Senthil2004}.

When considering only the $H^0_p$ system without $H_d$, there is a rigorous theorem originally proved by Mielke ~\cite{Mielke1991,Mielke1991a}, known as ``flat-band ferromagnetism''. It states that when an on-site Coulomb repulsion of any finite magnitude is turned on, the half-filled flat band has a fully-polarized FM ground state. It was later demonstrated that the flat-band ferromagnetism is stable against small perturbations. In some sense, the flat-band ferromagnetism represents the large-DOS limit of Stoner's criteria, complimentary to Nagaoka's ferromagnetism under the large-U limit.

Thus, with more $\pi$ electrons doped into the system, the flat-band ferromagnetism occurs and the itinerant fermions will act as a strong magnetic field on the spins via $H_{\pi d}$. Recall that the Hund's coupling $J_H$ is of the order of 100 meV, much larger than the AFM superexchange $J_d\simeq 7$ meV between spins. Consequently, the KAFM subsystem is also polarized into a FM configuration. Indeed, within the DFT+U formalism, we find that if we manually tunes the Fermi level to the half-filing point, which corresponds to extract 1/2 electron from each HAB on average, the spin degeneracy of the flat band is spontaneously lifted, and the FM Cu spin order has the lowest energy (Supporting Information).

It is straightforward to perform a spin-wave expansion on Eq.~\ref{eq:Heisenberg} with respect to the FM polarized ground state: $S_i^\dag=b_i, S_i^{-1}=b_i^\dag, S_i^z=\frac{1}{2}-b_i^\dag b_i$, in which $b$ corresponds to the Holstein-Primakoff boson, or magnon annihilation operator. Replacing the spin operators in Eq.~(\ref{eq:Heisenberg}) with $b^\dag$ and $b$, $H_d$ is transformed into:
\begin{eqnarray}
\label{eq:FM}
  H_d=\frac{J_d}{2}\sum_{\langle ij\rangle}(b_{i\sigma}^\dag b_{j\sigma} + \text{H.c.})+const.
\end{eqnarray}
This Hamiltonian appears almost the same as Eq.~(\ref{eq:Kagome}). The main difference is that the hopping particle transforms from a fermionic electron to a bosonic magnon. Obviously, the magnon band structure is the same as Fig.~\ref{fig:figure2}(b) after rescaling them with energy. It is important to note that the flat band is carried over to the magnon degree of freedom. The FM ground state is thus susceptible to various instabilities. In Fig.~\ref{fig:figure1}(c), we label this regime as a ``frustrated ferromagnet (FM)''.

\section*{Conclusion and perspective}
In conclusion, we have numerically demonstrated the coexistence of frustrated local spins and frustrated itinerant electrons in Cu$_3$(HAB)$_2$. Each of these two frustrated degrees of freedom is known to produce a rich phase diagram alone, and we argue that a real platform that simultaneously contains both will open even more possibilities. An effective model is constructed, but the theoretical solution has to rely on sophisticated numerical computations. Further experimental input is highly demanded, especially, the close energy proximity of the frustrated spin bands and charge bands enables a high tunability by gating to study the interplay between different quantum phases. One great challenge is to further improve the Cu$_3$(HAB)$_2$ crystalline quality. We have simulated the scanning tunneling spectroscopy and real-space mapping under different bias, which can be used to resolve the two types of frustrated lattices. The redox control techniques may also provide an alternate doping method to modify the electron filling of Cu$_3$(HAB)$_2$ lattice.

\matmethods{The crystal properties were calculated using first-principles methods based on density functional theory with Vienna \textit{ab} initio simulation package code. The molecular properties of HAB molecule and Cu-(HAB)$_2$ complex were calculated using Gaussian package with the B3LYP functional. More details are presented in $\textcolor{blue}{Supporting Information}$.

}

\showmatmethods 

\acknow{J.-W. Mei and F. Liu are supported by U.S. DOE-BES (Grant No. DE-FG02-04ER46148). W. J. is supported by the National Science Foundation-Material Research Science \& Engineering Center (NSF-MRSEC grand No. DMR-1121252). Z. L. is supported by NSFC under Grant No. 11774196. We thank the CHPC at the University of Utah and DOE-NERSC for providing the computing resources.}

\showacknow 



\bibliography{Tunable_bifrustration}{}
\bibliographystyle{plain}

\end{document}